\documentclass{article}
\usepackage{spconf,amsmath,amssymb, graphicx, algorithm, algpseudocode}

\title{Stochastic Vector Approximate Message Passing with applications to phase retrieval}

\name{Hajime Ueda, Shun Katakami, Masato Okada$^{\dag}$ \thanks{H.U. is supported by Advanced AI Talent Development to Lead the Next-Generation Intelligent Society (BOOST NAIS). This work is partly supported by JSPS KAKENHI Grant Number 23H00486, 23K16959, and Digital Transformation Initiative for Green Energy Materials (DX-GEM).  }}
\address{Graduate School of Frontier Sciences, The University of Tokyo, Kashiwa, Chiba
277-8561, Japan}

\begin{document}

\maketitle
\begin{abstract}
Phase retrieval refers to the problem of recovering a high-dimensional vector $\boldsymbol{x} \in \mathbb{C}^N$ from the magnitude of its linear transform $\boldsymbol{z} = A \boldsymbol{x}$, observed through a noisy channel. 
To improve the ill-posed nature of the inverse problem, it is a common practice to observe the magnitude of linear measurements $\boldsymbol{z}^{(1)} = A^{(1)} \boldsymbol{x},..., \boldsymbol{z}^{(L)} = A^{(L)}\boldsymbol{x}$ using multiple sensing matrices $A^{(1)},..., A^{(L)}$, with ptychographic imaging being a remarkable example of such strategies. 
Inspired by existing algorithms for ptychographic reconstruction, we introduce stochasticity to Vector Approximate Message Passing (VAMP), a computationally efficient algorithm applicable to a wide range of Bayesian inverse problems. By testing our approach in the setup of phase retrieval, we show the superior convergence speed of the proposed algorithm.
\end{abstract}


\section{Introduction}
\label{sec:intro}
\subsection{Background on Phase Retrieval}
\textit{Phase retrieval} is the problem of recovering a high-dimensional vector $\boldsymbol{x} \in \mathbb{C}^N$ from the observation $\boldsymbol{y} \in \mathbb{R}^M$ modeled by
\begin{align}
    \ \boldsymbol{z} = A\boldsymbol{x}\ , \ y_{\mu} \sim p_{\text{out}}(\ \cdot \  |\  |z_{\mu}|\ )  \ \ (\mu = 1,..., M)
\end{align}
where $A \in \mathbb{C}^{M \times N}$ is the sensing matrix and $p_{\text{out}}(\ \cdot \  |\  |z_{\mu}|\ )$ is a probability measure given the magnitude of the $\mu$-th entry of $\boldsymbol{z} \in \mathbb{C}^M$.
A typical application that fits this framework is \textit{Coherent Diffraction Imaging} (CDI) \cite{CDI_Review}, where an object image $\boldsymbol{x}$ is illuminated by coherent light, and the magnitude of diffracted light is observed in the far field.

It is of practical interest how many observations are needed to recover the signal $\boldsymbol{x}$.
In recent years, the information-theoretic threshold of sampling ratio has been intensively studied for phase retrieval with large random sensing matrices~\cite{Maillard2020}.
From an algorithmic viewpoint, spectral method ~\cite{Luo2018, Valzania2021} is known to be an efficient algorithm for obtaining an estimate $\hat{\boldsymbol{x}}$ correlated with $\boldsymbol{x}$, which is often employed to initialize iterative solvers, including the Wirtinger flow~\cite{WirtingerFlow}, and message passing algorithms~\cite{Mondelli2020ApproximateMP}.

In computational imaging, there are mainly two strategies to improve the ill-posed nature of phase reconstruction --  \textit{multiple sensing matrices} and \textit{exploitation of prior knowledge}.
A remarkable example of the former is \textit{ptychography} in CDI, where the magnitudes of several linear measurements $\boldsymbol{z}^{(1)} = A^{(1)} \boldsymbol{x},..., \boldsymbol{z}^{(L)} = A^{(L)}\boldsymbol{x}$ 
are observed:
In a typical setting of ptychography, each sensing matrix $A^{(l)}\in\mathbb{C}^{\bar{M}\times N} \ (l = 1,...,L)$ is given as $A^{(l)} = FPS^{(l)}$, where $F \in \mathbb{C}^{\bar{M}\times\bar{M}}$ is the 2D DFT, $P \in \mathbb{C}^{\bar{M}\times\bar{M}}$ is a diagonal matrix, and $S^{(l)} \in \{0,1\}^{\bar{M}\times N}$ is a binary matrix consisting of $\bar{M}$ row vectors of the identity matrix $I_N$.
On the forward model of ptychography, see, for example, \cite{Valzania2021, ADMM_ptychography}. Another example is \textit{coded diffraction pattern} \cite{CodedDiffractionPattern}, whose sensing matrices are expressed as $A^{(l)} = FP^{(l)}$, where $F \in \mathbb{C}^{N \times N}$ is the 2D DFT and $P^{(l)} = \text{Diag}(e^{j\theta^{(l)}_1}, \ldots, e^{j\theta^{(l)}_N})$ denotes a random phase modulation.
This model corresponds to CDI with "random phase masks" placed after the object $\boldsymbol{x}$, as illustrated in Fig.~1.
Random masks are actively utilized in CDI experiments \cite{coded_aparture_CDI}.

\begin{figure}[htb]
\begin{minipage}[b]{1.0\linewidth}
  \centering
  \centerline{\includegraphics[width=8.5cm]{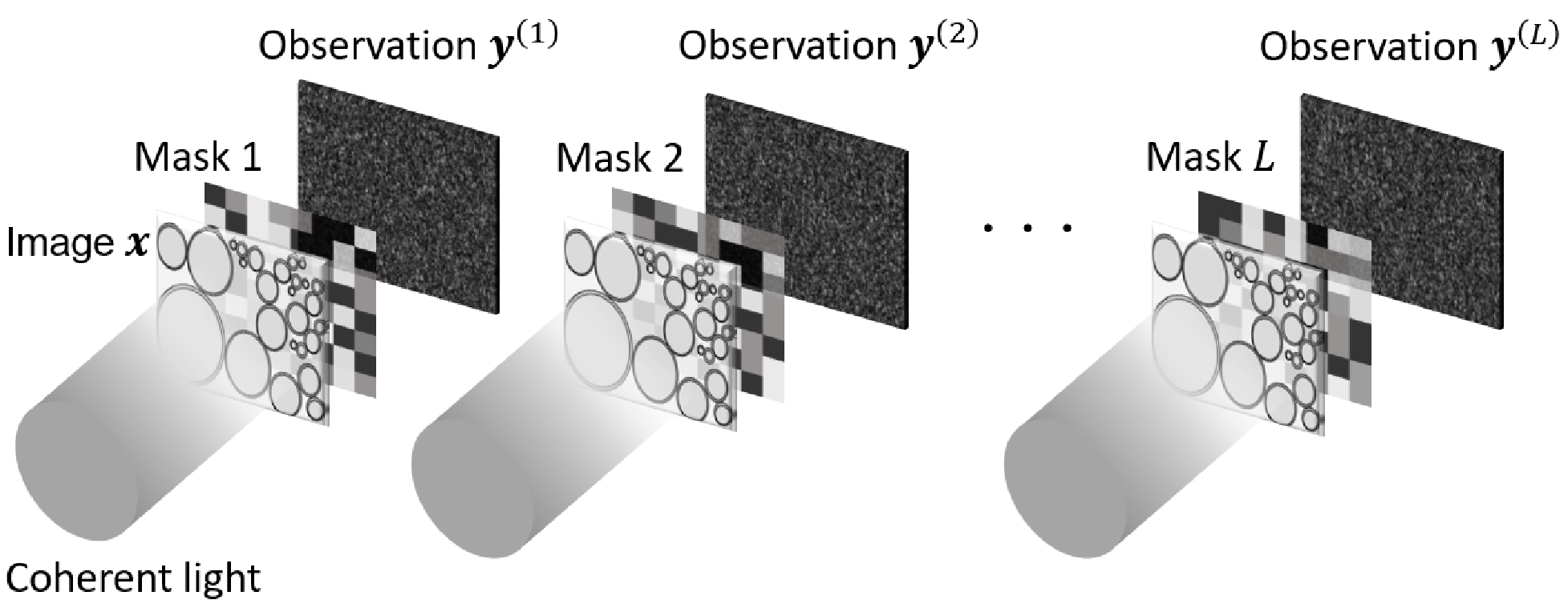}}
  \caption{Setup of coded diffraction pattern}\medskip
\end{minipage}
\end{figure}

Regarding the latter approach, \textit{Bayesian inference} provides a framework for incorporating prior knowledge, such as \textit{sparsity} and \textit{support constraint}, into a statistical model.
The Bayes' theorem for the model (1) is written as
\begin{align}
    p(\boldsymbol{x}, \boldsymbol{z} | \boldsymbol{y}) \propto p_{\text{in}}(\boldsymbol{x})  \delta (\boldsymbol{z} - A\boldsymbol{x})  \prod_{\mu = 1}^{M} p_{\text{out}} (y_{\mu} |  |z_{\mu}|) 
\end{align}
wherein the prior $p_{\text{in}}(\boldsymbol{x})$ expresses the prior knowledge on $\boldsymbol{x}$, and $\delta (\cdot)$ is the Dirac delta function. 
To compute the posterior density $p(\boldsymbol{x}, \boldsymbol{z} | \boldsymbol{y})$ in a reasonable computational time, approximations such as the mean-field approach and approximate message passing have been applied to phase retrieval~\cite{Maillard2020, PRVBEM, PRGAMP}.
These techniques can also be combined with deep neural network trained with natural images~\cite{deepEC}.

\subsection{Illustration of the key idea}
Our approach is to blend the two techniques stated above: multiple measurements and Bayesian inference. We aim to reconstruct a complex image $\boldsymbol{x} \in \mathbb{C}^N$ from $L$ observations $\boldsymbol{y}^{(l)} \in \mathbb{R}^{\bar{M}}$ $(l = 1, \ldots, L)$. Using sensing matrices $A^{(1)}, \ldots, A^{(L)} \in \mathbb{C}^{\bar{M} \times N}$, the observations are modeled by
\begin{align}
    \quad \boldsymbol{z}^{(l)} = A^{(l)} \boldsymbol{x}\ , \ \boldsymbol{y}^{(l)} \sim p_{\text{out}} (\cdot \mid |\boldsymbol{z}^{(l)}|)  \quad (l = 1,...,L)
\end{align}
with $p_{\text{out}} (\boldsymbol{y}\mid|\boldsymbol{z}|) \triangleq \prod_{\mu} p_{\text{out}} (y_{\mu} \mid |z_{\mu}|)$. By defining the concatenated sensing matrix and observation as
\begin{align}
    A = 
    \begin{pmatrix}
    A^{(1)} \\
    \vdots \\
    A^{(L)}
    \end{pmatrix}, \ 
    \boldsymbol{z} = 
    \begin{pmatrix}
    \boldsymbol{z}^{(1)} \\
    \vdots \\
    \boldsymbol{z}^{(L)}
    \end{pmatrix}, \ 
    \boldsymbol{y} = 
    \begin{pmatrix}
    \boldsymbol{y}^{(1)} \\
    \vdots \\
    \boldsymbol{y}^{(L)}
    \end{pmatrix}
\end{align}
the model (3) is reduced to model (1). However, crucial to our algorithm is sequential access to the  observations $\boldsymbol{y}^{(1)}, \ldots, \boldsymbol{y}^{(L)}$ which leads to faster convergence, compared to accessing all of the observations at once.
Indeed, sequential ptychographic solvers such as ePIE~\cite{ePIE} is known to effectively avoid bad local minima, as detailed in ~\cite{LSsolver}.

To incorporate multiple measurements into a Bayesian framework, we extend Vector Approximate Message Passing (VAMP) ~\cite{VAMP, GVAMP}, an efficient inference algorithm for generalized linear models.
Our extension, coined Stochastic VAMP, allows "mini-batch learning" commonly employed in ptychographic reconstruction\cite{LSsolver} and speeds up convergence. 
This method is analogous to Stochastic Gradient Descent in non-convex optimization, where gradients are computed from data subsets, hence the name Stochastic VAMP.

\section{Stochastic VAMP}
\label{sec:method}
\subsection{Derivation of VAMP}
We review the derivation of VAMP based on expectation propagation (EP) ~\cite{VAMP, GVAMP}. For the Bayesian problem in Eq. (1), we start with an equivalent model: 
\begin{align} 
p(\boldsymbol{x}_1, \boldsymbol{x}_2, \boldsymbol{z}_1, \boldsymbol{z}_2 | \boldsymbol{y}) \propto\  &p_{\text{in}}(\boldsymbol{x}_1) \delta(\boldsymbol{x}_1 - \boldsymbol{x}_2) \times\\\nonumber
 & \delta (A\boldsymbol{x}_2 - \boldsymbol{z}_2)  \delta(\boldsymbol{z}_1 - \boldsymbol{z}_2) p_{\text{out}}( \boldsymbol{y}| |\boldsymbol{z}_1|)
\end{align}
The corresponding factor graph is shown in Fig.2 (a). In belief propagation (BP) \cite{BP}, probability distributions called \textit{messages} are passed through the graph to form \textit{beliefs}, which are the estimates of marginal distributions of each variable.
EP approximates messages by using Gaussian. Applying EP to the graph in Fig.2 (a) yields the VAMP algorithm.

\begin{figure}[htb]
\begin{minipage}[b]{1.0\linewidth}
  \centering
  \centerline{\includegraphics[width=8.5cm]{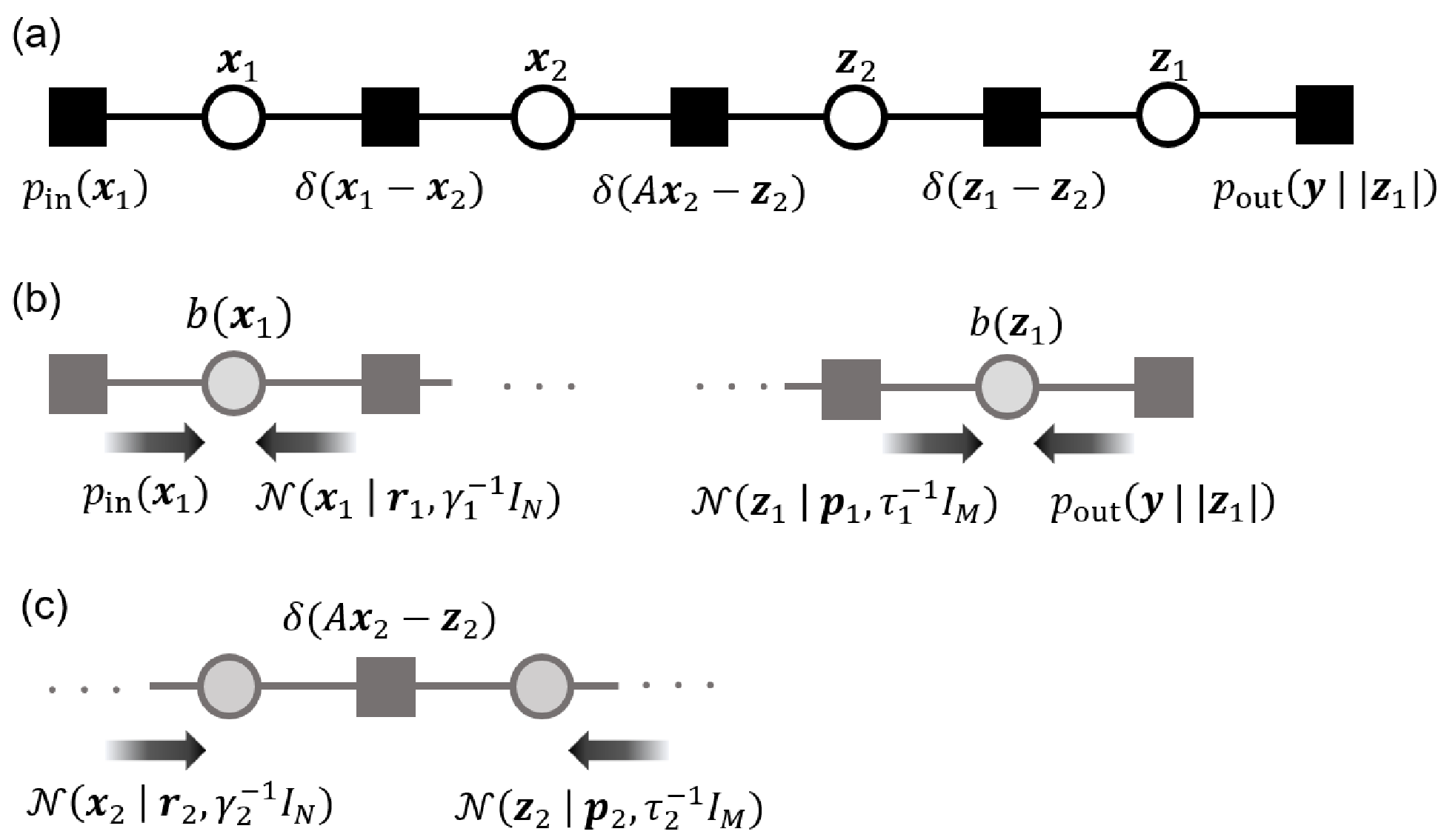}}
  \caption{(a) Factor graph (b) Denoising (c) LMMSE estimation}\medskip
\end{minipage}
\end{figure}

Here, we outline the update rules of VAMP. Each iteration includes two steps: \textit{denoising} and \textit{LMMSE estimation}.
In the denoising part, $\boldsymbol{x}_1$ node inputs the message $\mathcal{N}(\boldsymbol{x}_1|\boldsymbol{r}_1, \gamma_1^{-1} I_N)$ from $\delta(\boldsymbol{x}_1 - \boldsymbol{x}_2)$ node.\footnote{The density of the complex circular Gaussian distribution is given as $\mathcal{N}(\boldsymbol{x} | \boldsymbol{r}, \gamma^{-1} I_N ) \propto \exp (- \gamma ||\boldsymbol{x} - \boldsymbol{r}||_2^2)$, wherein $\boldsymbol{x}, \boldsymbol{r} \in \mathbb{C}^N$ and  $\gamma > 0$.}
This creates the belief $
    b(\boldsymbol{x}_1) \propto p_{\text{in}}(\boldsymbol{x}_1) \mathcal{N}(\boldsymbol{x}_1|\boldsymbol{r}_1, \gamma_1^{-1} I_N)
$.
Denoting expectation w.r.t. measure $b(\boldsymbol{x}_1)$ by $\mathbb{E}[\ \cdot\  | b(\boldsymbol{x}_1)]$, the denoiser functions are defined as
\begin{align}
    \boldsymbol{g}_x(\boldsymbol{r}_1, \gamma_1) &= \mathbb{E}[\boldsymbol{x}_1 | b(\boldsymbol{x}_1)]\\
    \boldsymbol{g}'_x(\boldsymbol{r}_1, \gamma_1) &= \frac{1}{N} \mathbb{E}\left[ || \boldsymbol{x}_1 - \boldsymbol{g}_x(\boldsymbol{r}_1, \gamma_1) ||_2^2 | b(\boldsymbol{x}_1)\right]
\end{align}
$b(\boldsymbol{x}_1)$ is approximated by $\mathcal{N}(\boldsymbol{x}_1 | \hat{\boldsymbol{x}}_1, \eta_1^{-1} I_N)$ with 
\begin{align}
    \hat{\boldsymbol{x}}_1 = \boldsymbol{g}_x(\boldsymbol{r}_1, \gamma_1)\ ,\  \eta_1^{-1} = \boldsymbol{g}'_x(\boldsymbol{r}_1, \gamma_1)
\end{align}
Then, the message $\mathcal{N}(\boldsymbol{x}_2 | \boldsymbol{r}_2, \gamma_2^{-1} I_N)$ is sent to $\boldsymbol{x}_2$ node from $\delta(\boldsymbol{x}_1 - \boldsymbol{x}_2)$ node, which, by the rules of BP, is given by
\begin{align}
    \gamma_2 = \eta_1 - \gamma_1 \ , \  \boldsymbol{r}_2 = (\eta_1 \hat{\boldsymbol{x}}_1 - \gamma_1 \boldsymbol{r}_1 )/ \gamma_2
\end{align}
Likewise, the message $\mathcal{N}(\boldsymbol{z}_1 | \boldsymbol{p}_1, \tau_1^{-1} I_M)$ is input to $\boldsymbol{z}_1$ node.
$\hat{\boldsymbol{z}}_1, \lambda_1, \boldsymbol{p}_2$, and $\tau_2$ are give by
\begin{align}
  &\boldsymbol{g}_z(\boldsymbol{p}_1, \tau_1) = \mathbb{E}[\boldsymbol{z}_1 | b(\boldsymbol{z}_1)]\\
  &\boldsymbol{g}'_z(\boldsymbol{p}_1, \tau_1) = \frac{1}{M} \mathbb{E}\left[ || \boldsymbol{z}_1 - \boldsymbol{g}_z(\boldsymbol{p}_1, \tau_1) ||_2^2 | b(\boldsymbol{z}_1)\right]\\
  &\hat{\boldsymbol{z}}_1 = \boldsymbol{g}_z(\boldsymbol{p}_1, \tau_1)\ ,\  \lambda_1^{-1} = \boldsymbol{g}'_z(\boldsymbol{p}_1, \tau_1)\\
  &\tau_2 = \lambda_1 - \tau_1 \ , \  \boldsymbol{p}_2 = (\lambda_1 \hat{\boldsymbol{z}}_1 - \tau_1 \boldsymbol{p}_1 )/ \tau_2
\end{align}
with $b(\boldsymbol{z}_1) \propto \mathcal{N}(\boldsymbol{z}_1 | \boldsymbol{p}_1, \tau_1^{-1} I_M) p_{\text{out}}(\boldsymbol{y}| |\boldsymbol{z}_1|)$.

In the LMMSE estimation, joint belief on $\boldsymbol{x}_2$ and $\boldsymbol{z}_2$ is given as  $
    b(\boldsymbol{x}_2, \boldsymbol{z}_2) \propto \mathcal{N}(\boldsymbol{x}_2 | \boldsymbol{r}_2, \gamma_2^{-1} I_N) \times \delta(A\boldsymbol{x}_2 - \boldsymbol{z}_2) \mathcal{N}(\boldsymbol{z}_2 | \boldsymbol{p}_2, \tau_2^{-1} I_M)
$.
Marginalizing $b(\boldsymbol{x}_2, \boldsymbol{z}_2)$ w.r.t. $\boldsymbol{z}_2$, we have
$
    b(\boldsymbol{x}_2) \propto \mathcal{N}(\boldsymbol{x}_2 | \boldsymbol{r}_2, \gamma_2^{-1} I_N) \times \mathcal{N}(A\boldsymbol{x}_2 | \boldsymbol{p}_2, \tau_2^{-1} I_M)
$.
This is then approximated by $\mathcal{N}(\boldsymbol{x}_2 | \hat{\boldsymbol{x}}_2, \eta_2^{-1} I_N)$ with 
\begin{align}
    &\hat{\boldsymbol{x}}_2 = Q(\gamma_2 \boldsymbol{r}_2 + \tau_2 A^H \boldsymbol{p}_2)\ ,\  \eta_2^{-1} = \frac{1}{N} \text{Tr} (Q) \\
    &Q = (\gamma_2 I_N + \tau_2 A^H A)^{-1}
\end{align}
In the rest of this paper, we focus on the case where $A^H A$ is diagonal, thus the matrix inversion in Eq. (15) is not the dominant computational cost. For general sensing matrices, per-iteration matrix inversion can be avoided by pre-computing the singular value decomposition of $A$.
Similarly, $b(\boldsymbol{z}_2)$ is approximated by $\mathcal{N}(\boldsymbol{z}_2 | \hat{\boldsymbol{z}_2}, \lambda_2^{-1} I_M)$ with
\begin{align}
    \hat{\boldsymbol{z}}_2 = A\hat{\boldsymbol{x}}_2\ , \lambda_2^{-1} = \frac{1}{M} \text{Tr}(AQA^H)
\end{align}
The messages $\mathcal{N}(\boldsymbol{x}_1|\boldsymbol{r}_1, \gamma_1^{-1} I_N)$ and $\mathcal{N}(\boldsymbol{z}_1 | \boldsymbol{p}, \tau_1^{-1} I_M)$ are updated according to
\begin{align}
    &\gamma_1 = \eta_2 - \gamma_2\ , \boldsymbol{r}_1 = (\eta_2 \hat{\boldsymbol{x}}_2 - \gamma_2 \boldsymbol{r}_2)/\gamma_1 \\
    &\tau_1 = \lambda_2 - \tau_2\ , \boldsymbol{p}_1 = (\lambda_2 \hat{\boldsymbol{z}}_2 - \tau_2 \boldsymbol{p}_2)/\tau_1
\end{align}
which closes the VAMP iteration.

\subsection{Stochastic VAMP}
A natural extension of VAMP to multiple measurements is derived by EP on the factor graph in Fig.3. 
Denote the message sent from $\delta (\boldsymbol{z}^{(l)}_1 - \boldsymbol{z}^{(l)}_2)$ to $\boldsymbol{z}^{(l)}_1$ and the message sent from $\delta (\boldsymbol{z}^{(l)}_1 - \boldsymbol{z}^{(l)}_2)$ to $\boldsymbol{z}^{(l)}_2$ by $\mathcal{N}\left(\boldsymbol{z}^{(l)}_1 | \boldsymbol{p}^{(l)}_1, (\tau^{(l)}_1)^{-1} I_M\right)$ and $\mathcal{N}\left(\boldsymbol{z}^{(l)}_2 | \boldsymbol{p}^{(l)}_2, (\tau^{(l)}_2)^{-1} I_M\right)$, respectively.
It is straight forward to derive the LMMSE estimator to update $(\boldsymbol{p}^{(l)}_1,\tau^{(l)}_1)$ and $(\boldsymbol{p}^{(l)}_2,\tau^{(l)}_2)$ for $l = 1,...,L$.
Note that if we update these \( L \) messages in parallel, the resulting algorithm is essentially the same as VAMP with the sensing matrix \( A \) and observation \( \boldsymbol{y} \) defined in Eq. (4). The only difference from VAMP is that we allow \( \{\tau^{(l)}_1, \tau^{(l)}_2\}\) \((l = 1,...,L)\) to take different values. However, we can also update \( L \) messages sequentially, yelding Stochastic VAMP specified in Algorithm 1.

\begin{figure}[htb]
\begin{minipage}[b]{1.0\linewidth}
  \centering
  \centerline{\includegraphics[width=8.5cm]{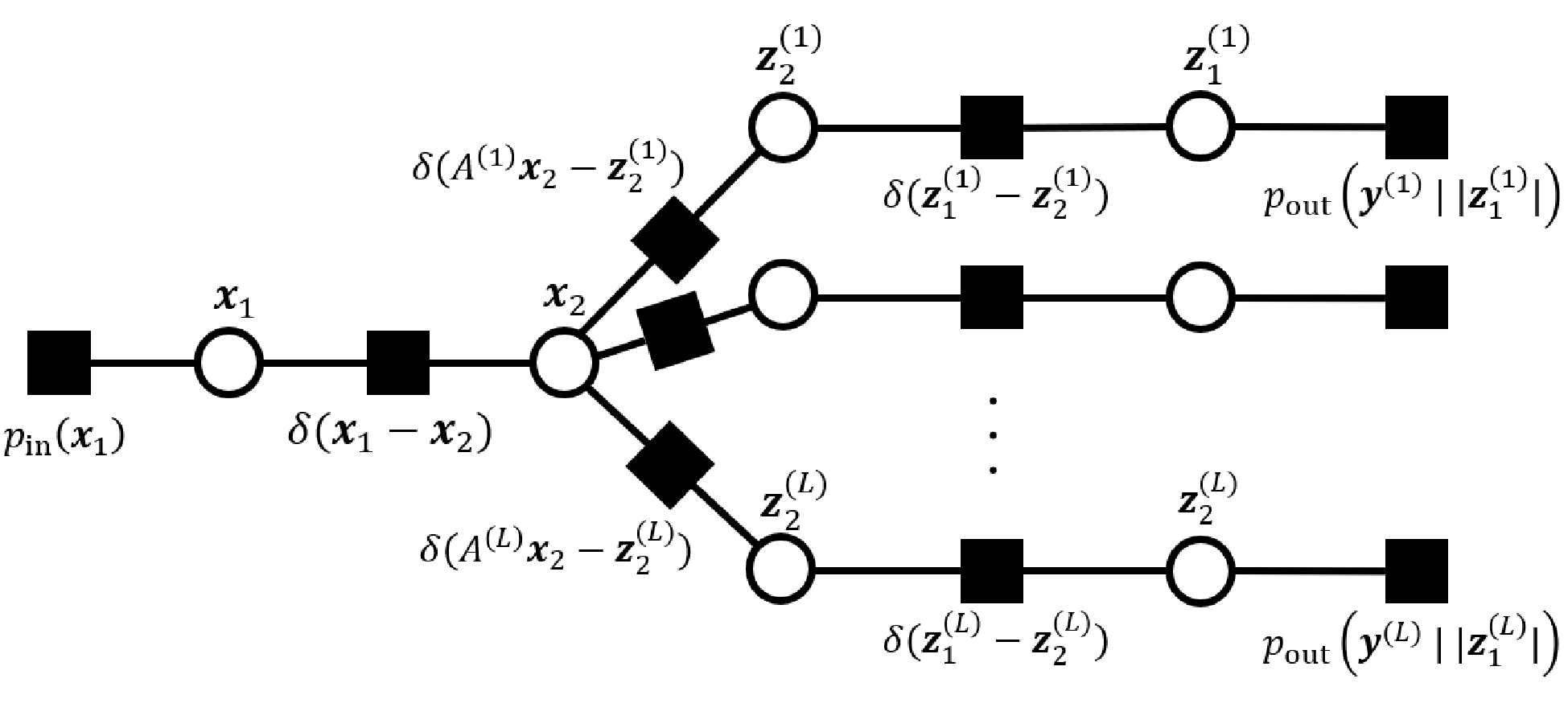}}
  \caption{Factor graph of stochastic VAMP}
\end{minipage}
\end{figure}

\begin{algorithm}[t]
  \caption{Stochastic VAMP}\label{alg:cap}
  \begin{algorithmic}[1]
  \Require Sensing matrices $A^{(1)},..., A^{(L)}$, observations $\boldsymbol{y}^{(1)},...,\boldsymbol{y}^{(L)}$, number of iterations $K_{\text{it}}$
  \State Initialize $\boldsymbol{r}_{2}, \gamma_{2} > 0$ and $\boldsymbol{p}_{2}, \tau_{2} > 0\ \ (l = 1,...,L)$
  \For{$k = 0,1,...,K_{\text{it}}$}
  \For{$l = 1,...,L$}
  \State //LMMSE estimation
  \State $Q = \left(\gamma_{2} I_N + \sum_{l=1}^L \tau^{(l)}_{2} (A^{(l)})^H A^{(l)}\right)^{-1}$
  \State $\hat{\boldsymbol{x}}_{2} = Q\left(\gamma_{2} \boldsymbol{r}_{2} + \sum_{l=1}^L \tau^{(l)}_{2} (A^{(l)})^H \boldsymbol{p}^{(l)}_{2}\right)$
  \State $\eta_{2}^{-1} = \frac{1}{N} \text{Tr}(Q)$
  \State $\hat{\boldsymbol{z}}_{2} = A^{(l)} \hat{\boldsymbol{x}}_{2}$
  \State $(\lambda^{(l)}_{2})^{-1} = \frac{1}{M} \text{Tr}\left(A^{(l)} Q (A^{(l)})^H\right)$
  \State // Message passing
  \State $\gamma_{1} = \eta_{2} - \gamma_{2}$
  \State $\boldsymbol{r}_{1} = (\eta_{2} \hat{\boldsymbol{x}}_{2} - \gamma_{2} \boldsymbol{r}_{2})/\gamma_{1}$
  \State $\tau_{1}^{(l)} = \lambda_{2}^{(l)} - \tau_{2}^{(l)}$
  \State $\boldsymbol{p}^{(l)}_{1} = (\lambda_{2}^{(l)} \hat{\boldsymbol{z}}_{2} - \tau_{2}^{(l)} \boldsymbol{p}^{(l)}_{2})/\tau_{1}^{(l)}$
  \State //Denoising
  \State $\hat{\boldsymbol{x}}_{1} = \boldsymbol{g}_x (\boldsymbol{r}_{1}, \gamma_{1})$
  \State $\eta_{1}^{-1} = \boldsymbol{g}'_x (\boldsymbol{r}_{1}, \gamma_{1})$
  \State $\hat{\boldsymbol{z}}^{(l)}_{1} = \boldsymbol{g}_z (\boldsymbol{p}^{(l)}_{1}, \tau_{1})$
  \State $\lambda_{1}^{-1} = \boldsymbol{g}'_z (\boldsymbol{p}^{(l)}_{1}, \tau_{1})$
  \State // Message passing
  \State $\gamma_{2} = \eta_{1} - \gamma_{1}$
  \State $\boldsymbol{r}_{2} = (\eta_{1} \hat{\boldsymbol{x}}_{1} - \gamma_{1} \boldsymbol{r}_{1})/\gamma_{2}$
  \State $\tau_{2}^{(l)} = \lambda_{1}^{(l)} - \tau_{1}^{(l)}$
  \State $\boldsymbol{p}^{(l)}_{2} = (\lambda_{1}^{(l)} \hat{\boldsymbol{z}}_{1} - \tau_{1}^{(l)} \boldsymbol{p}^{(l)}_{1})/\tau_{2}^{(l)}$
  \EndFor
  \EndFor
  
  \State Return $\hat{\boldsymbol{x}}_{1}$
  \end{algorithmic}
  \end{algorithm}

\subsection{Algorithmic detail}
It is standard to stabilize the convergence of VAMP by employing the "damping" method, where $\boldsymbol{r}_1$ in Eq. (17) is treated as "raw" value $\boldsymbol{r}_{\text{raw}}$, and $\boldsymbol{r}_1$ is updated as $\boldsymbol{r}_1 = \rho \boldsymbol{r}_{\text{raw}} + (1 - \rho) \boldsymbol{r}_{\text{old}}$, in which $0 < \rho < 1$ and $ \boldsymbol{r}_{\text{old}}$ is the previous value of $\boldsymbol{r}_1$.
Similar technique is applied to the update of $\gamma, \boldsymbol{p}, $ and $\tau$.  
Notably, Stochastic VAMP requires less damping than standard VAMP, contributing to a faster convergence.

When each matrix $A^{(l)}$ have $(A^{(l)})^H A^{(l)}$ diagonal, the matrix inversion in line 5 adds only minor complexity to the algorithm.
The dominant computational cost comes from the matrix-vector multiplications in lines 6 and 8.
In the case of coded diffraction patterns, the cost is $2L$ FFTs per iteration.
For general sensing matrices, costly matrix-inversion can still be avoided by using the graphical model introduced for D-VAMP~\cite{DVAMP} and pre-computing the singular value decompositions of $A^{(1)},...,A^{(L)}$.

In our experiments, we assumed that the SNR and prior are known, but they can be learned from the dataset using methods from ~\cite{EMVAMP}.
Although noise learning and multiple restarts can help phase retrieval algorithms avoid bad local minima~\cite{PRVBEM, PRGAMP}, numerical results indicate that Stochastic VAMP remains stable even without these enhancements.
Python codes for the numerical experiments in this paper is available at https://github.com/sacbow/StochasticVAMP.

\section{Computational results}
\label{sec:simulation}
\subsection{Common setup}
In this section, we consider an observation model defined as:
\begin{align}
    \boldsymbol{y} = |\boldsymbol{z} + \boldsymbol{w}|
\end{align}
where $\boldsymbol{w}$ is the Gaussian noise given as $\boldsymbol{w} \sim \mathcal{N}(\cdot | \boldsymbol{0}, \gamma_w^{-1} I_M)$, and $|\cdot|$ is the element-wise absolute value.
The corresponding likelihood function $p_{\text{out}}(\boldsymbol{y} | |\boldsymbol{z}|)$ is refered to as the \textit{Rician distribution}, whose denoising functions are derived in ~\cite{PRGAMP}. We assume a Gaussian prior $p_{\text{in}}(\boldsymbol{x}) = \mathcal{N} (\boldsymbol{x} | \boldsymbol{0}, I_N)$, but structured priors can also be used.
The total number of observed values is $M = L\bar{M}$, where $L$ is the number of sensing matrices $A^{(1)},..., A^{(L)}$ and $A^{(l)} \in \mathbb{C}^{\bar{M} \times N}$.
The sampling ratio is $\alpha \triangleq \frac{M}{N}$. 
Since the posterior density is invariant w.r.t. global phase shift, the deviation between the true signal $\boldsymbol{x}$ and VAMP estimate $\hat{\boldsymbol{x}}$ is measured by 
\begin{align}
  \text{NMSE} \triangleq \underset{\theta \in [0, 2\pi)} {\operatorname{min}} \frac{||\boldsymbol{x} - e^{j\theta} \hat{\boldsymbol{x}}||^2}{||\boldsymbol{x}||^2}
\end{align}

\subsection{Haar distributed sensing matrix}
We consider sensing matrices $A^{(l)}$ with $(A^{(l)})^H A^{(l)} = I_{N}$ drawn from Haar distribution.
In the case of noiseless phase retrieval, the threshold value of $\alpha$ above which VAMP recovers the true signal $\boldsymbol{x}$ is investigated via asymptotic analysis~\cite{Maillard2020}, in which the threshold value for Haar-distributed $A$ is estimated as $\alpha_{\text{FR, Alg}} \simeq 2.265$.
In our experiment, we used $N = 512$, $\alpha = 2.4$, and $L = 2$, with SNR of 30 dB.
We compared Stochastic VAMP with vanilla VAMP, which uses concatenated sensing matrices and observations (Eq.(4)).
While VAMP required damping of $\rho = 0.9$ for convergence, Stochastic VAMP was stable with less amount of damping ($\rho = 0.97$).
The true signal $\boldsymbol{x}$ was drawn from the prior, and both algorithms were randomly initialized.
The complexity of both algorithms is similar, with $O((L-1)N)$ difference per-iteration.
The NMSE is shown against iteration number $k$ in Fig. 4.
We observed that the stochasticity significantly accelerates the convergence of the phase retrieval algorithm.

\begin{figure}[htb]
\begin{minipage}[b]{1.0\linewidth}
  \centering
  \centerline{\includegraphics[width=8.5cm]{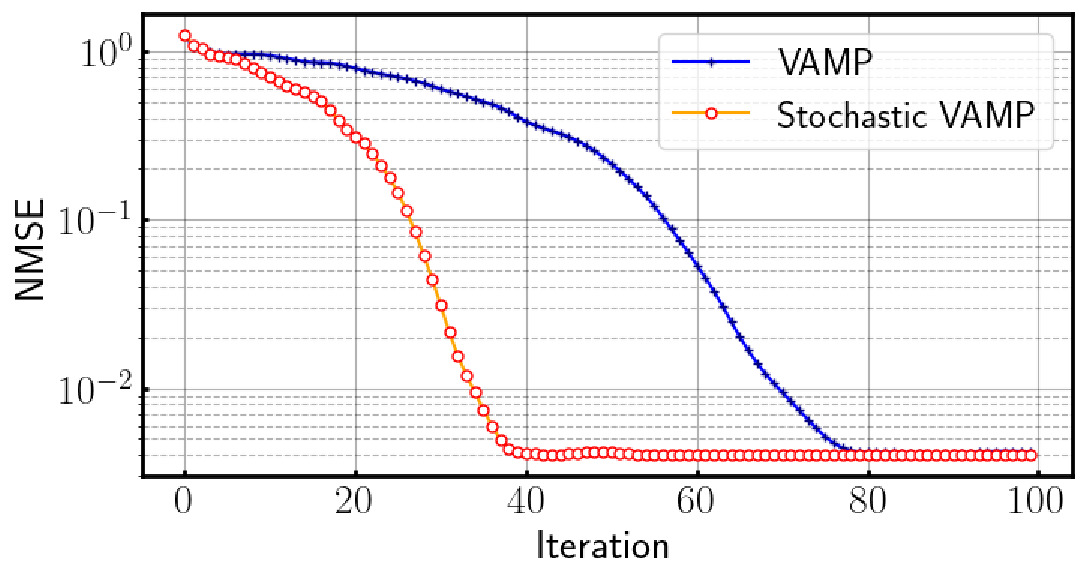}}
  \caption{NMSE versus iteration number for VAMP and Stochastic VAMP.}\medskip
\end{minipage}
\end{figure}

\subsection{Coded diffraction pattern from real world image}
In this section, we evaluate the performance of Stochastic VAMP in a more realistic setting.
Using the sensing matrix for coded diffraction patterns with $L = 3$ random masks, we conducted numerical experiments on a $256 \times 256$-pixel real-valued image (Fig.5(a)). 
The SNR is set to 30 dB.
A typical reconstruction is shown in Fig.5(b), which was obtained within 20 iterations with a computational time of only \textbf{3.9} seconds.
Notably, this rapid convergence is impressive, given that conventional phase retrieval algorithms typically require $10^2$ to $10^3$ iterations for reliable reconstruction.
Although we assumed complex Gaussian prior, the reconstruction can be further improved by using non-negative real prior.

\begin{figure}[htb]
\begin{minipage}[b]{1.0\linewidth}
  \centering
  \centerline{\includegraphics[width=8.5cm]{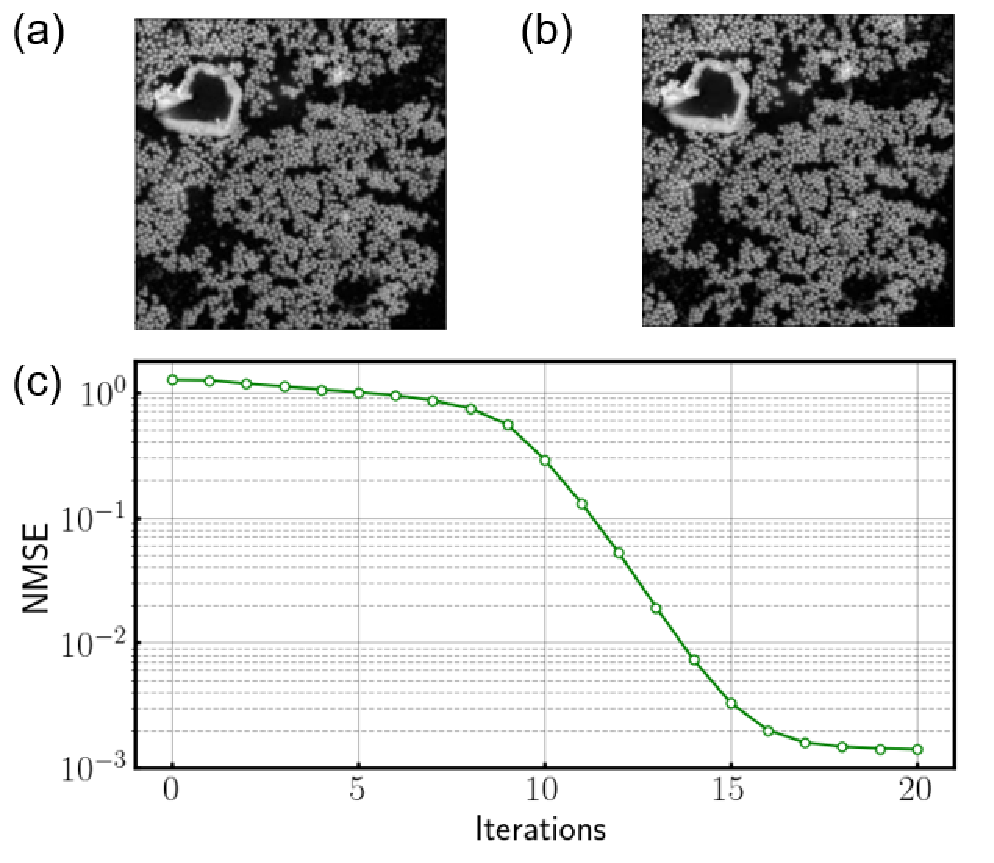}}
  \caption{(a) test image (b) reconstructed image (c) NMSE versus iteration number for Stochastic VAMP. }\medskip
\end{minipage}
\end{figure}

\section{Conclusions}
\label{sec:conclusion}
We enhanced VAMP by introducing stochasticity through non-parallel message updates in belief propagation.
In phase retrieval with multiple measurements, our approach significantly accelerates convergence while maintaining the per-iteration complexity of conventional algorithms.

Being a Bayesian approach, our method allows for incorporating prior knowledge, such as image sparsity, similar to ~\cite{PRGAMP}.
Future work includes applying this approach to ptychographic imaging.
Practical ptychographic phase retriecal algorithms such as ePIE perform self-calibration by which both the probe and the object image are reconstructed.
Probe reconstruction could be achieved by integrating Stochastic VAMP with techniques from ~\cite{BiVAMP, BiGVAMP}.

\vfill\pagebreak

\bibliographystyle{IEEEbib}
\bibliography{refs}
\end{document}